\documentclass[conference]{IEEEtran}
\IEEEoverridecommandlockouts
\usepackage{cite}
\usepackage{amsmath,amssymb,amsfonts}
\usepackage{graphicx}
\usepackage{textcomp}
\usepackage{xcolor}
\usepackage{svg}
\usepackage{mathrsfs}
\usepackage[ruled,linesnumbered,commentsnumbered]{algorithm2e}
\usepackage{epstopdf}
\def\BibTeX{{\rm B\kern-.05em{\sc i\kern-.025em b}\kern-.08em
    T\kern-.1667em\lower.7ex\hbox{E}\kern-.125emX}}
\begin{document}

\title{A Novel Chaos Theory Inspired Neuronal Architecture }
%
\author{\IEEEauthorblockN{ Harikrishnan N B and Nithin Nagaraj}
\IEEEauthorblockA{\textit{Consciousness Studies Programme} \\
\textit{National Institute of Advanced Studies}\\
\textit{Indian Institute of Science Campus, Bengaluru, India}\\
harikrishnannb07@gmail.com, nithin.nagaraj@gmail.com}

}

\maketitle

\begin{abstract}
The practical success of widely used machine learning (ML) and deep learning (DL) algorithms in Artificial Intelligence (AI) community owes to availability of large datasets for training and huge computational resources. Despite the enormous practical success of AI, these algorithms are only loosely inspired from the biological brain and do not mimic any of the fundamental properties of neurons in the brain, one such property being the chaotic firing of biological neurons. This motivates us to develop a novel neuronal architecture where the individual neurons are intrinsically chaotic in nature. By making use of the \emph{topological transitivity} property of chaos, our neuronal network is able to perform classification tasks with very less number of training samples. For the MNIST dataset\footnotemark \footnotetext{http://yann.lecun.com/exdb/mnist/index.html}, with as low as $0.1 \%$ of the total training data, our method outperforms ML and matches DL in classification accuracy for up to $7$ training samples/class. For the Iris dataset\footnotemark \footnotetext{http://archive.ics.uci.edu/ml/datasets/iris}, our accuracy is comparable with ML algorithms, and even with just two training samples/class, we report an accuracy as high as $95.8 \%$. This work highlights the effectiveness of chaos and its properties for learning and paves the way for chaos-inspired neuronal architectures by closely mimicking the chaotic nature of neurons in the brain. 
\end{abstract}

\section{Introduction}
Next to the universe, the human brain is the most complex and sparsely understood system. Brain science is said to be in \emph{Faraday stage}\cite{ramachandran1998phantoms}, which means our current understanding of the working of the brain is at a very primitive stage.  It has been estimated that human brain has approximately 86 billion neurons \cite{86_billion_neurons} which interact with each other forming a very complex system. Neurons are inherently nonlinear and found to exhibit chaos\cite{therechaos1}, \cite{therechaos2}. An interesting counter-intuitive property of networks of neurons in the brain is their ability to learn in the presence of enormous amount of noise and neural interference~\cite{snr_neuron_2015measuring}. Inspired by the biological brain, researchers have developed Artificial Intelligent systems which use learning algorithms such as Deep Learning (DL) and Machine Learning (ML) that loosely mimic the human brain.


DL and ML algorithms have a wide variety of practical applications in computer vision, natural language processing, speech processing\cite{speechrnn}, cyber-security~\cite{harikrishnan2018machine}, medical diagnosis~\cite{ding2018deep} etc. However, these algorithms do not use the essential properties of human brain. One such property of brain is the presence of chaotic neurons \cite{therechaos1}, \cite{therechaos2}. Even though Artificial Neural Networks are biologically inspired, none of its varied architectures have neurons which exhibit chaos though it has been shown that certain type of neural networks exhibit chaotic dynamics (for e.g., in Recurrent Neural Networks~\cite{chaotic_rnn}). Chaotic regimes with a wide range of behaviors are beneficial for the brain to quickly adapt to changing conditions~\cite{therechaos1}. There has also been some evidence that weak chaos is good for learning~\cite{sprott2013chaos}. Inspired by these studies, in this work, we explore whether chaos can be useful in learning algorithms.

There have been previous attempts at developing novel biologically inspired learning architectures. A recent research study by~\cite{delahunt2018putting} proposes a learning architecture that uses a mathematical model of the olfactory network of moth to train to read MNIST\cite{lecun-mnisthandwrittendigit-2010}. The same study~\cite{delahunt2018putting} also highlights learning from limited data samples. In another interesting research~\cite{kathpalia2019novel}, the authors propose a novel compression based neural architecture for memory encoding and decoding that uses a 1D chaotic map known as Generalized Lur\"{o}th Series (GLS)~\cite{nagaraj2008novel}. GLS coding, a generalized form of Arithmetic coding \cite{nagaraj2009arithmetic}, is used for memory encoding in their work.

In this work, we propose for the first time - a novel neuronal architecture of GLS neurons and train it for a classification task using a unique property of chaos known as \emph{Topological Transitivity} (TT). This research is a first step towards building a more realistic brain-inspired learning architecture. Here, chaos is used at the level of individual neurons. As we shall demonstrate, one of the key benefits of our proposed method is its superior performance in low training samples regime. 

The paper is organized as follows. We introduce the novel architecture in Section II along with a detailed description of the topological transitivity based classification algorithm. This is followed by experiments and results in section III. We conclude by highlighting the unique advantages of our method while also mentioning some of the possible future research directions in section IV.




\begin{figure*}
    \centerline{ \includegraphics[width=0.5\textwidth]{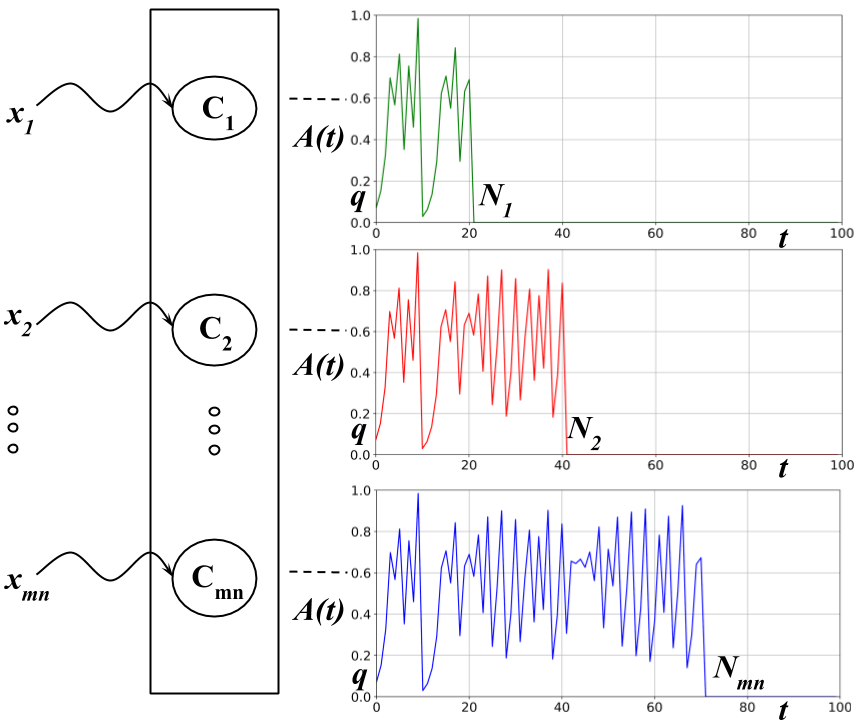}}
    
    \caption{The proposed Chaotic GLS neural network architecture. $C_1, C_2,..., C_{mn}$ represent the input layer of chaotic GLS neurons each with an initial normalized membrane potential of $q$ units. $\{ x_i\}_{1}^{mn}$ is the normalized set of stimuli that is input to the network. Each GLS neuron fires (chaotically) until its membrane potential $A(t)$ is in the neighbourhood of the stimulus. The {\it firing time} $N_i$ $ms$ of the corresponding GLS neuron $C_i$ is the {\it topological transitivity} based extracted feature.}
    \label{model_archi}
    \end{figure*}

\section{The Proposed Architecture}
The basic diagram of the proposed neural architecture is provided in Figure~\ref{model_archi}. It comprises of an input layer of GLS neurons represented as $C_1, C_2,..., C_{mn}$. The GLS neuron is a 1D chaotic map which we shall describe shortly. The GLS neurons get activated in the presence of a stimulus (input data) which results in a chaotic firing pattern. Each GLS neuron in the input layer continues to fire chaotically until its amplitude matches that of the stimulus - at which point it stops firing. In the model provided in Figure~\ref{model_archi}, $x_1, x_2,.., x_{mn}$ represents the stimulus (normalized) which is assumed to be a real number between $0$ and $1$. Each GLS neuron has an initial normalized membrane potential of $q$ units (a real number between $0$ and $1$) which is also the initial value of the chaotic map. In general, each GLS neuron can have a different initial normalized membrane potential though in this work we assume that they are all the same. The GLS neurons have a refractory period of 1 millisecond which means that the inter-firing interval is 1 $ms$ (from the instant they are presented with a stimulus). When the GLS neuron encounters a stimulus say $x_k$, the neuron starts firing chaotically until it matches the amplitude of the stimulus, i.e., when the normalized membrane potential reaches a neighbourhood of $x_k$, at which time it stops firing. The time duration $N_k$ $ms$ for which the $k$-th GLS neuron is active is defined as the {\it firing time}. The firing of each GLS neuron is guaranteed to halt (as soon as its membrane potential reaches the neighbourhood of $x_k$) owing to the property of Topological Transitivity $(TT)$ which is defined below. 

\textbf{Definition 1} {\it Topological Transitivity:} Given a map $T: Q \rightarrow Q$, $T$ is said to be topologically transitive on $Q$, if for every pair of non-empty open sets $U$ and $V$ in $Q$, there exist a non negative integer $n$ and a real number $u \in U$ such that $T^{n}(u) \in V$.

In our example, $T$ is the 1D GLS chaotic map with $Q: [0,1)$. We define $U: (q-\epsilon, q+\epsilon)$ and $V: (x_k-\epsilon, x_k+\epsilon)$ as the two non-empty open sets and $\epsilon > 0$. It follows from the above definition that there exists $N_k \geq 0$ (integer) and a real number $u \in U$ such that $T^{N_k}(u) \in V$. We take $u=q$. It may be the case that certain values of $q$ may not work, but we can always find a $q$ that works.

\subsection{GLS Neuron: Chaotic map}
\begin{figure}[htbp]
\centerline{ \includegraphics[width=0.25\textwidth]{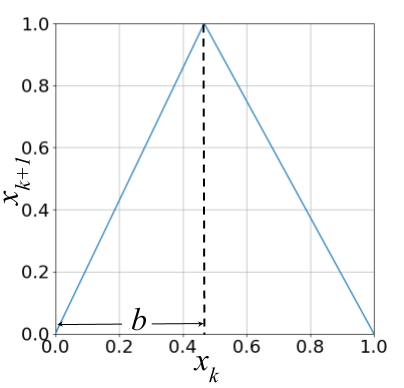}}

\caption{The first return map of Generalized Lur\"{o}th Series (GLS)~\cite{dajani2002ergodic, nagaraj2009arithmetic}. GLS is a chaotic 1D map that exhibits {\it topological transitivity}. }
\label{map}
\end{figure}

The GLS neuron~\cite{dajani2002ergodic} is a 1D map $T: [0,1) \rightarrow [0,1)$ defined as: 
\\
$T(x)  =  \left\{\begin{matrix}
\frac{x}{b}&, ~~~~ 0 \leq x < b, \\ 
\frac{(1-x)}{(1 - b)}&, ~~~~ b \leq x < 1 
\end{matrix}\right. $\\
where $x \in [0,1)$. We have chosen $b = 0.467354$ in our study. Figure~\ref{map} represents the GLS map. 

\subsection{Topological Transitivity (TT) based classification algorithm}
Let \emph{\textbf{X}} be a $m \times n$ matrix where each row represents distinct data instance and the columns represent the different features. Each row (data instance) of \emph{\textbf{X}} is  $x^{i} = [x_{1}^i, x_{2}^i, x_{3}^i,..., x_{n}^i ]$ $ \in $ ${\rm I\!R}^{n}$. If the data instances are images then each $x^{i}$ represents a vectorized image with each $x_k^i$ representing a pixel value. In our study, we have normalized\footnote{For a non-constant matrix $X$, normalization is achieved by performing $\frac{X-\min(X)}{\max(X)-\min(X)}$. A constant matrix $X$ is normalized to all ones.} the values of the matrix \emph{\textbf{X}} to lie in $[0,1]$. 

There are mainly three steps in TT based classification algorithm.

\begin{itemize}
    \item \textbf{TT based feature extraction} - Algorithm 2 represents the TT based feature extraction. 
    
    Let $I_{k}^{i} = (x_{k}^{i}-\epsilon, x_{k}^{i}+\epsilon)$ be the $\epsilon$-neighbourhood of $x_{k}^{i}$ where $\epsilon > 0$. Let $N_{k}^{i}$ be the {\it firing time} of the $ik$-th GLS Neuron when subjected to the normalized stimulus $x_{k}^{i}$. This is nothing but the time in $ms$ or equivalently the number of iterations of the GLS map $T$ that is required to reach the interval $I_{k}^{i}$ starting from the initial membrane potential $q$. In other words, $q \rightarrow T_{k}(q) \rightarrow T_{k}^{2}(q) \rightarrow T_{k}^{3}(q)... \rightarrow T_{k}^{N_i}(q) $ where $T_{k}^{N_i}(q) \in I_{k}^{i}$ for the first time. The GLS neuron stops firing as soon as this is satisfied and we shall call $N_{k}^{i}$ as {\it TT based feature}.

    \item \textbf{Training} - Algorithm 1 represents the TT based training step.
    Let us assume there are $s$ classes $\mathscr{C}_1$, $\mathscr{C}_2$, \ldots, $\mathscr{C}_s$ with labels $1, 2, \ldots, s$ respectively. Let the data belonging to $\mathscr{C}_1$, $\mathscr{C}_2$, \ldots, $\mathscr{C}_s$  be denoted as $X^1, X^2, \ldots,  X^s$ respectively. For simplicity, let us assume that $X^1, X^2, \ldots,  X^s$ are $s$ distinct matrices of size $m \times n$. The TT based feature extraction step is applied to $X^1, X^2, \ldots, X^s$ separately to yield $Y^{1}, Y^{2}, \ldots, Y^{s}$.  $Y^{1}, Y^{2}, \ldots, Y^{s}$ have the same size as $X^{1}, X^{2}, \ldots, X^{s}$ since TT based feature extraction is applied to each stimulus. The average across rows is computed as follows:   
\begin{eqnarray*}
    M^{1} &=& \frac{1}{m} \Big[ \sum_{i = 1}^{m} Y^{1}_{i1}, \sum_{i = 1}^{m} Y^{1}_{i2}, \ldots, \sum_{i = 1}^{m} Y^{1}_{in} \Big], \\
M^{2} &=& \frac{1}{m} \Big[ \sum_{i = 1}^{m} Y^{2}_{i1}, \sum_{i = 1}^{m} Y^{2}_{i2}, \ldots, \sum_{i = 1}^{m} Y^{2}_{in} \Big],\\
&\vdots&\\
M^{s} &=& \frac{1}{m} \Big[ \sum_{i = 1}^{m} Y^{s}_{i1}, \sum_{i = 1}^{m} Y^{s}_{i2}, \ldots, \sum_{i = 1}^{m} Y^{s}_{in} \Big].
\end{eqnarray*}
These $s$ row-vectors $M^1, M^2, \ldots, M^s$ are termed as {\it representation} vectors for the $s$ classes. Each vector $M^k$ is the {\it average internal representation} of all the stimuli corresponding to class $k$. This is biologically analogous to internal representations of experiences induced by storing {\it memory traces} corresponding to distinct classes in the brain.


\end{itemize}

\begin{itemize}
    \item \textbf{Testing} - Algorithm 3 represents the testing part. Let the normalized test data be an $r \times n$ matrix denoted as $Z$. The $i$th test data instance of $Z$ is denoted as $z^i = [z^{i}_{1}, z^{i}_{2}, z^{i}_{3}..., z^{i}_{n}]$. TT based feature extraction is performed for each of the test data instances $(z^i)$ where $i = 1, 2, 3, \ldots, r$. Let the resultant TT based feature extracted matrix be denoted as $F$, where $f^{i} =[f^{i}_1, f^{i}_2, \ldots, f^{i}_n]$ is the $i$th row of $F$. Now we compute the cosine similarity of $f^{i}$  individually with each of the {\it representation} vectors $M^1, M^2,..., M^s$ as follows:

\begin{eqnarray*}
\cos(\theta_{1}) &=& \frac{f^{i} \cdot M^{1}}{\left \|f^{i} \right \|_2 \left \|M^{1} \right \|_2}, \\
\cos(\theta_{2}) &=& \frac{f^{i} \cdot M^{2}}{\left \|f^{i} \right \|_2 \left \|M^{2} \right \|_2},\\
&\vdots&\\
\cos(\theta_{s}) &=& \frac{f^{i} \cdot M^{s}}{\left \|f^{i} \right \|_2 \left \|M^{s} \right \|_2},
\end{eqnarray*}
where $\left \|v \right \|_2$ is the $l_2$ norm of row-vector $v$ and $f^i \cdot M^j$ is the dot product between the row-vectors $f^i$ and $M^j$. From the above $s$ scalar values we find that index ($p$) which corresponds to the maximum cosine similarity between the vector $M^p$ and $f^i$:
\begin{equation*}
      \theta_p  = \arg\max_{\theta_i}  (\cos(\theta_{1}), \cos(\theta_{2}), \ldots, \cos(\theta_{s})).
\end{equation*}
%
%
Now, the index $p$ is assigned as the class label for $z^{i}$. This step is repeated until each instance of the test data is assigned a class label. 
\end{itemize}
\begin{figure}[htbp]
\centerline{ \includegraphics[width=0.45\textwidth]{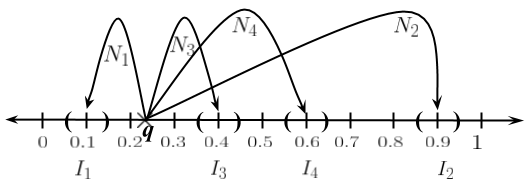}}

\caption{Illustration of topological transitivity based feature extraction (Algorithm 2) for an example. Starting from the initial normalized membrane potential of $q$ units, it takes $N_i$ iterations to reach the neighbourhood $I_i$ of the $i$-th stimulus.}
\label{TTillustration_4intervals}
\end{figure}

{\it Example: }We explain the aforementioned steps with the help of an example. For simplicity, let us assume a binary classification problem with two classes $\mathscr{C}_1$ and $\mathscr{C}_2$ with class labels $1$ and $2$ respectively. Let the input data be a  matrix $X = \big[\frac{X^1}{X^2}\big]$ of size $4 \times 4$ where $X^1$ represents the first two rows of $X$ which is the data belonging to class-$1$ and $X^2$ represents the remaining two rows of $X$ which is the data belonging to class-$2$. The input layer of the proposed neuronal architecture (Figure~\ref{model_archi}) consists of $16$ neurons which are denoted as $C_1, C_2, \ldots, C_{16}$. The initial membrane potential for each of these neurons is set to $q = 0.23$ units. As an example, consider the first row of $X^1$: $\hat{x}^{1} = [1, 9, 4, 6]$. The first step is to normalize the data. After normalization, let us say we have  $x^1= [ 0.1, 0.9, 0.4, 0.6]$. These are the stimulus to the input layer of the GLS neural network (for the first four GLS neurons: $C_1, C_2, C_3$ and $C_4$). The stimulus initiates the firing of these 4 GLS neurons. Let us assume the {\it firing times} are $N_1, N_2, N_3, N_4$ milliseconds. As depicted in Figure~\ref{TTillustration_4intervals}, it takes $N_k$ number of iterations to reach  $I_k = (x^{1}_k - \epsilon, x^{1}_k + \epsilon)$ which is the neighbourhood of the $k$-th stimulus. Choosing $\epsilon=0.05$, the four neighbourhoods are  $I_1 = (0.05, 0.15), I_2 = (0.85, 0.95), I_3 = (0.35, 0.45), I_4 = (0.55, 0.65)$. Similary, $N_5, N_6, \ldots, N_{16}$ are the {\it firing times} for the GLS neurons from $C_5, C_6, \dots, C_{16}$ respectively. This completes the TT based feature extraction step.



\begin{figure}[htbp]
\centerline{ \includegraphics[width=0.5\textwidth]{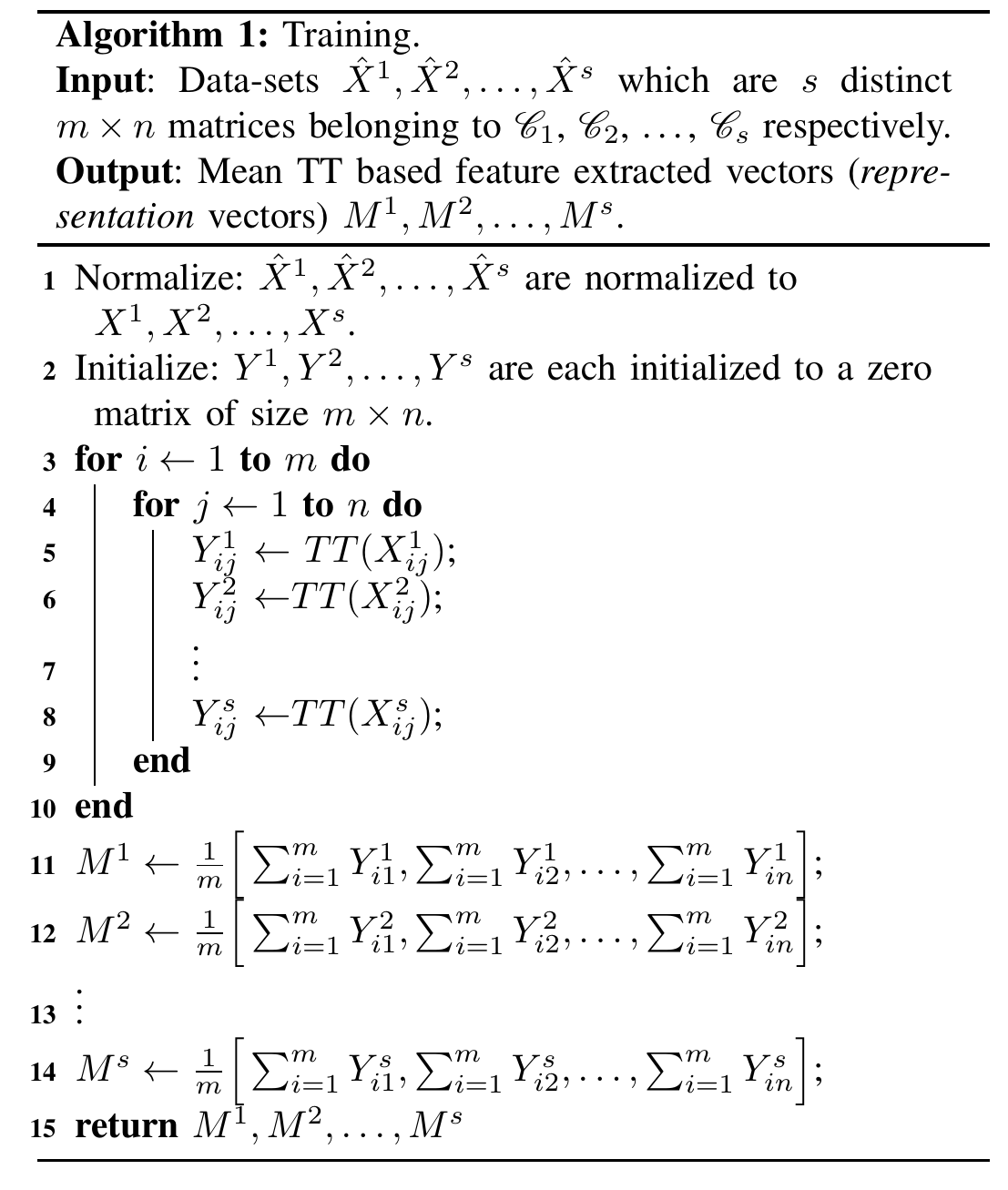}}

\label{alg:train}
\end{figure}

%
%

At the beginning of the training step, the TT based features extracted from the data are arranged as: 
$
Y^1 = 
\begin{bmatrix}  
N_1&N_2&N_3&N_4\\
N_5&N_6&N_7&N_8\\
\end{bmatrix},
Y^2 = 
\begin{bmatrix}  
N_9&N_{10}&N_{11}&N_{12}\\
N_{13}&N_{14}&N_{15}&N_{16}\\
\end{bmatrix}.
$
In the training step, we compute the two {\it representation} vectors corresponding to the two classes as $M^1 = \frac{1}{2}[N_1+N_5, N_2+N_6, N_3+N_7, N_4+N_8 ]$ and $M^2 = \frac{1}{2}[N_9+N_{13}, N_{10}+N_{14}, N_{11}+N_{15}, N_{12}+N_{16} ]$. 

Once the {\it representation} vectors are computed, we are ready to perform the testing on unseen data. Assume that the test data is a matrix $Z$ of size $2 \times 4$. We are required to classify each row of $Z$ to belong to either of class  $\mathscr{C}_1$ or $\mathscr{C}_2$. We first normalize the matrix $Z$ so that it contains only real values between 0 and 1. The TT based features are extracted for $Z$ by recording the firing times of the GLS neurons to yield
$
F = 
\begin{bmatrix}  
f^{1}_1&f^{1}_2&f^{1}_3&f^{1}_4\\
f^{2}_5&f^{2}_6&f^{2}_7&f^{2}_8\\
\end{bmatrix}
$. In order to classify $f^1$, the first row of $F$ (and hence the first row of $Z$), we compute the cosine similarity measure between $f^1$ and the two representation vectors $M^1$ and $M^2$ independently (say $\cos(\theta_1)$ and $\cos(\theta_2)$). We find the maximum of these two values, say $\cos(\theta_2)$. In this case, the label assigned to the first row of $Z$ would be 2. We repeat this for the second row of $F$. In this way, the unseen test data is classified using the representation vectors.  
%
%
%
\begin{figure}[htbp]
\centerline{ \includegraphics[width=0.5\textwidth]{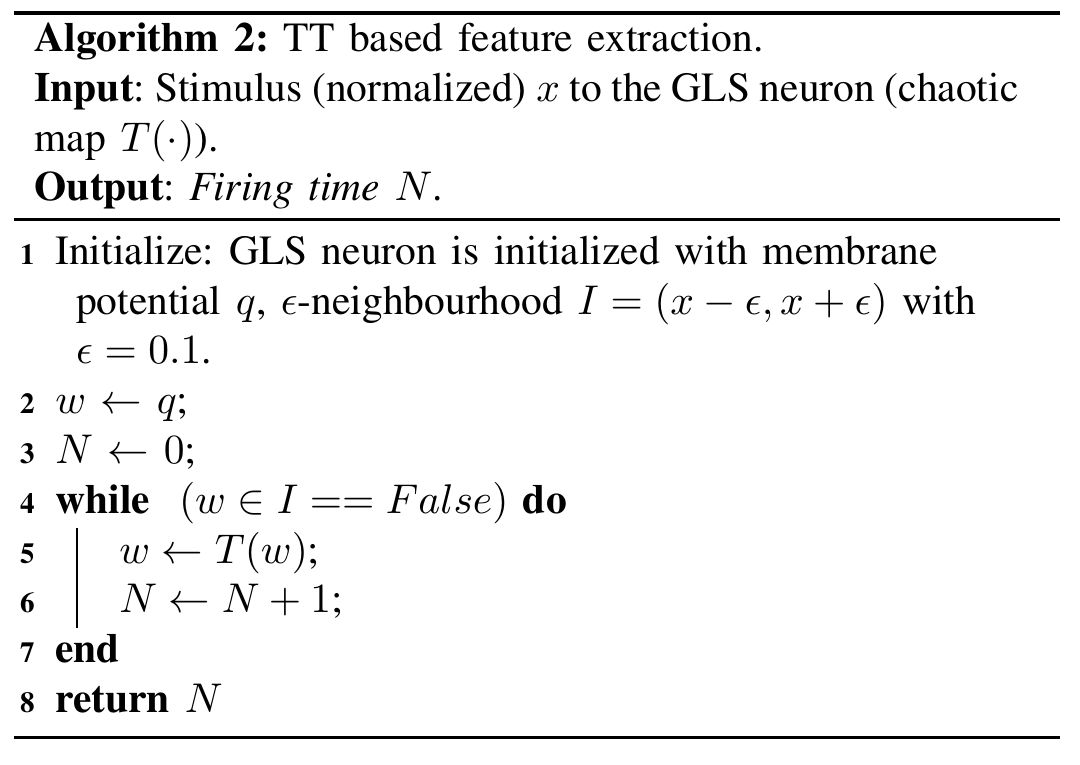}}
\label{alg:TT}
\centerline{ \includegraphics[width=0.5\textwidth]{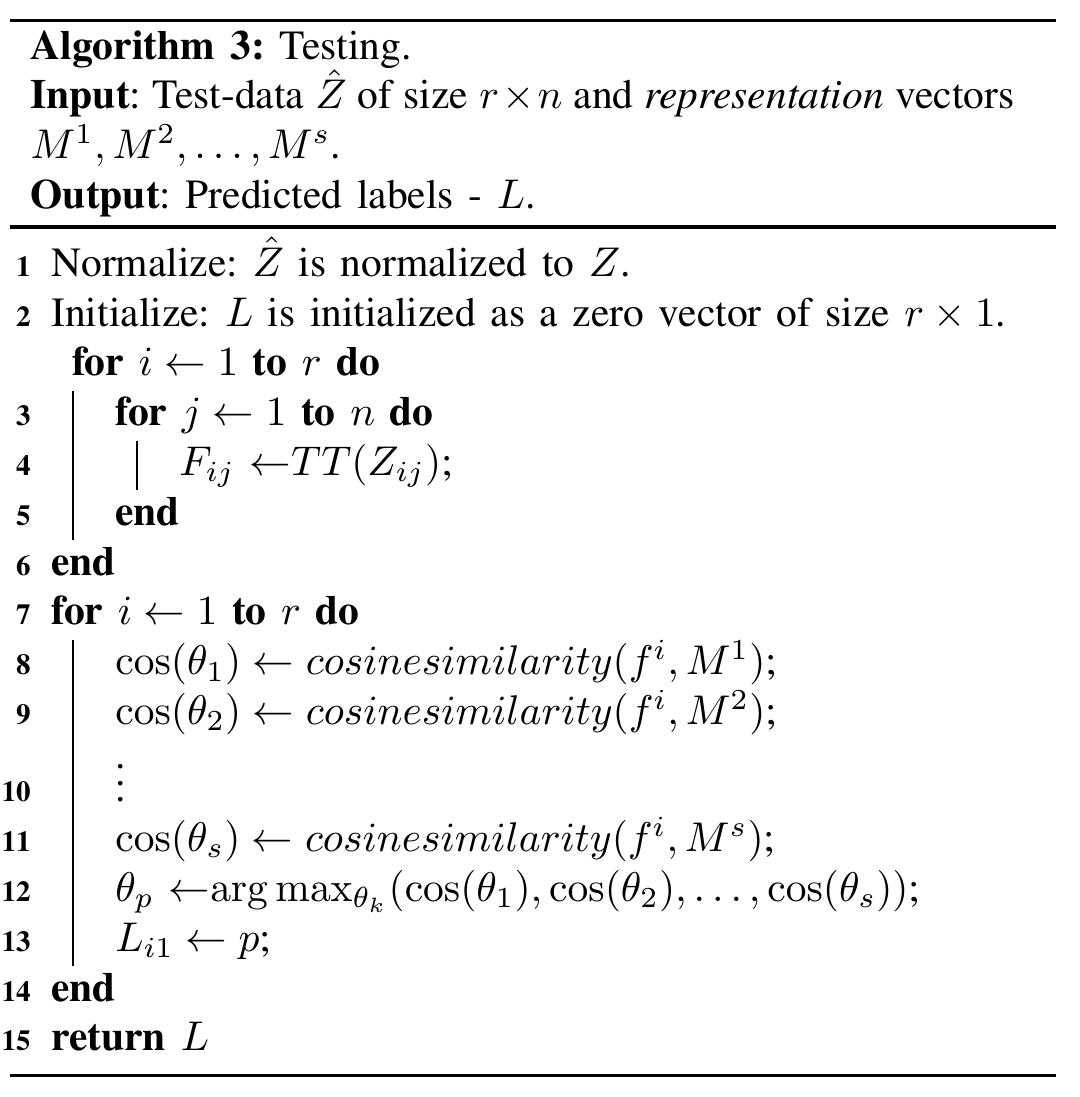}}
\label{alg:test}
\centerline{ \includegraphics[width=0.5\textwidth]{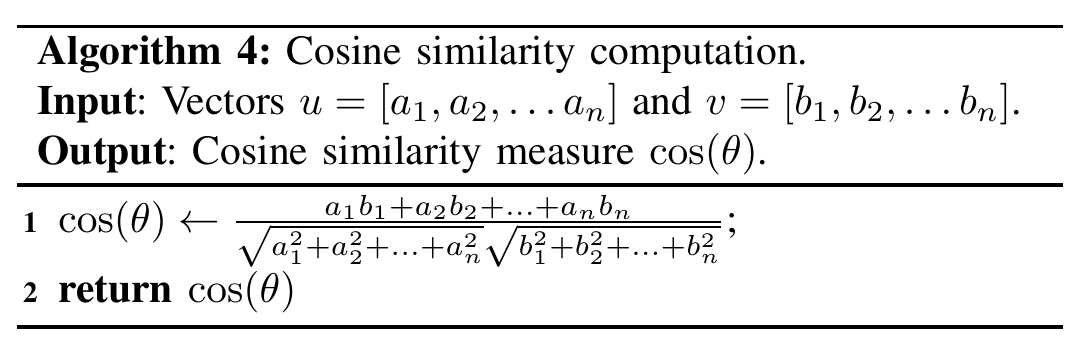}}
\label{alg:cosinesimilarity}
\end{figure}
\subsection{Hyperparameters} 
The hyperparameters used in this method are as follows:
\begin{enumerate}
    \item Map and its properties: In the proposed algorithm, we used 1D GLS chaotic map for the neuron. In general, we can also use other chaotic maps (such as logistic map) that satisfy the topological transitivity property. In the GLS map used in the proposed method (Figure~\ref{map}), $b$ is another hyperparameter. 
    \item The  initial  normalized  membrane  potential $q$, which is also the initial value for the chaotic map, is another hyperparameter. This initial value can be different for each GLS neuron, though in our work we have chosen the same for all the neurons. 
\end{enumerate}
The above hyperparameters can be tuned to further improve the performance. 
\section{Experiments and Results}
Learning from limited samples is a challenging problem in the AI community.  We evaluate the performance of our proposed TT based classification on MNIST~\cite{lecun-mnisthandwrittendigit-2010} and Iris data~\cite{blake1998uci} with limited training data samples. A brief description of these datasets is given below.

\subsection{Datasets}

\subsubsection{MNIST}
This dataset consists of hand written digits from 0 to 9 stored as digitized 8-bit grayscale images with dimensions of $28$ pixels $\times$ $28$ pixels with a total of $60000$ images for training and $10000$ images for testing. This is a 10-class classification task, i.e., the goal is to automatically classify these images to the ten classes corresponding to the digits 0 to 9. In our study, we performed independent trials of training with only $1, 2, 3, \dots, 20, 21$ data samples per class (randomly chosen from the available $60000$ training images). For each trial, we tested our algorithm with $10000$ unseen test images.
\subsubsection{Iris data}
This dataset consists of 4 attributes of 3 types (classes) of Iris plants (namely Setosa, Versicolour and Virginica). The 4 attributes are: sepal length (cm), sepal width (cm), petal length (cm) and petal width (cm). There are $50$ data samples per class. This is a $3$-class classification problem. In this study, we performed independent trials of training with randomly chosen $1, 2, 3, \ldots, 5$ data samples per class. For each trail, we tested with $120$ unseen randomly chosen data samples. 
%




\subsection{Comparative performance evaluation of the proposed method with other methods}
We compare our method with existing algorithms in literature. Specifically, we compare with Decision Tree (DT), Support Vector Machine (SVM), K-Nearest Neighbour (KNN) and 2-layer neural network. The machine learning techniques used in this research are implemented using {\it Scikit-learn}~\cite{scikit-learn}. The default parameters in {\it Scikit-learn} for DT, SVM and KNN are used in this research. We have used {\it gini} criterion for DT classifier and {\it radial basis function} (RBF) kernel for SVM based classification. For KNN, the number of nearest neighbours used was $5$. We have used {\it Keras}~\cite{chollet2015keras} package for the implementation of $2$-layer neural network with $784$ neurons in the input layer, $784$ neurons in the hidden layer and $10$ neurons in the output layer for MNIST classification task. For Iris data classification task, we used $4$ neurons in the input layer, $4$ neurons in the hidden layer and $3$ neurons in the output layer.

\begin{figure}[h!]
    \centerline{ \includegraphics[width=0.45\textwidth]{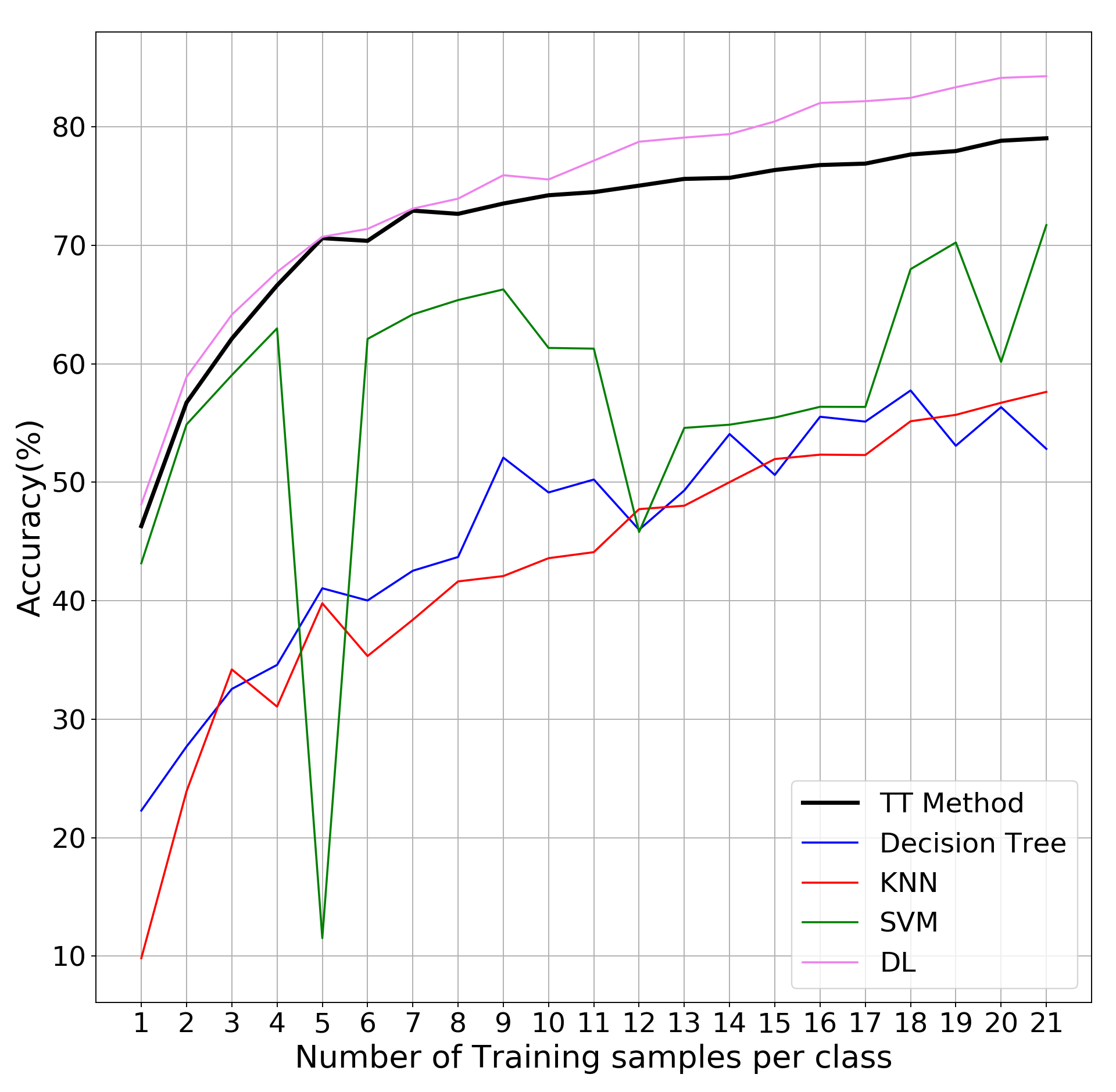}}
    
    \caption{Comparative performance evaluation of TT based method with DT, SVM, KNN and DL ($2$-layers) for MNIST dataset in the low training sample regime. }
    \label{m1vsML_mnist}
    \end{figure}

\begin{figure}[h!]
    \centerline{ \includegraphics[width=0.45\textwidth]{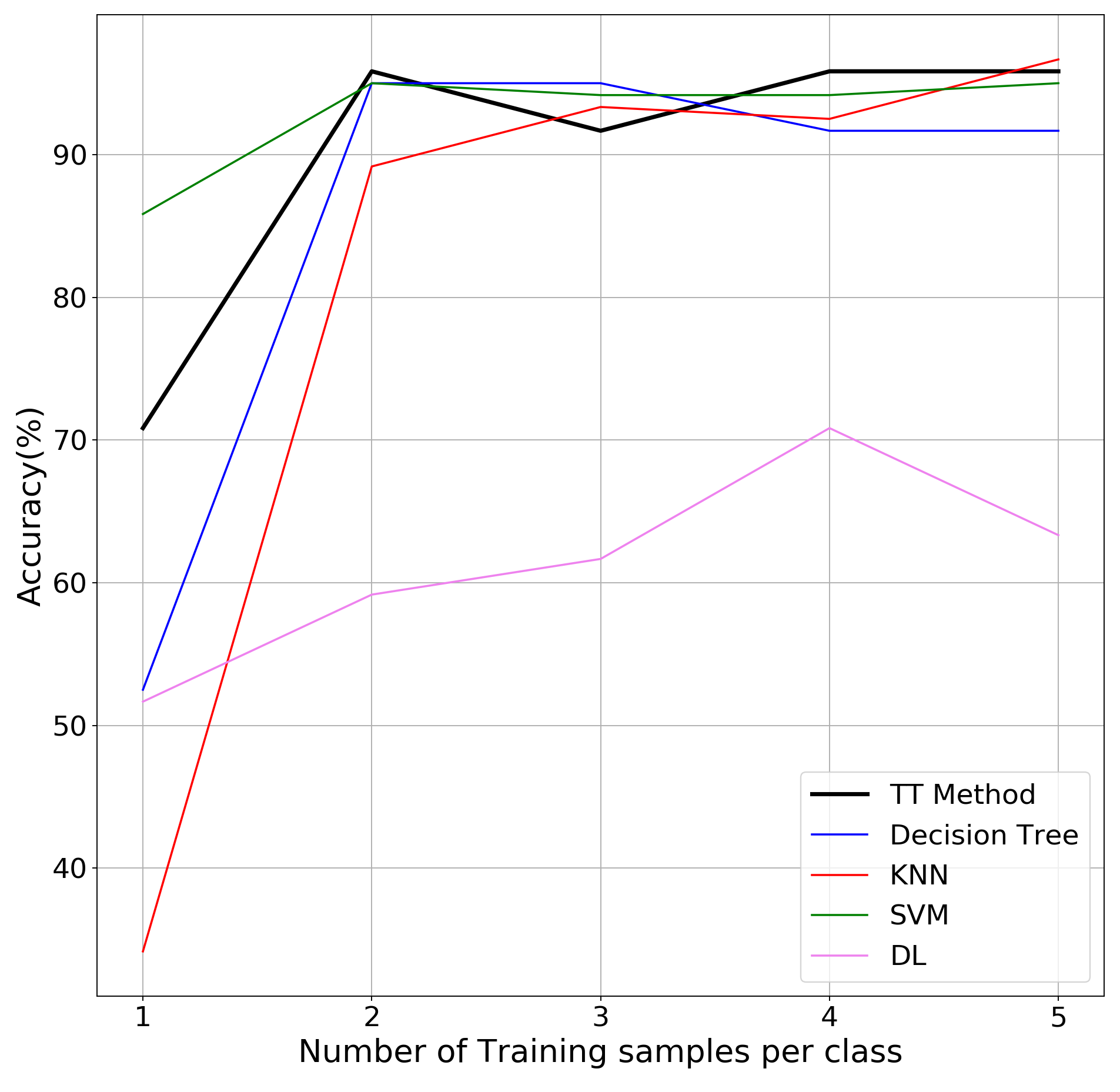}}
    
    \caption{Comparative performance Comparison of TT based method with DT, SVM, KNN and DL ($2$-layers) for Iris dataset in the low training sample regime.}
    \label{m1vsML_iris}
    \end{figure}

The comparative performance of TT based method and ML methods for MNIST and Iris data are provided in Figure~\ref{m1vsML_mnist} and~\ref{m1vsML_iris} respectively. From these results, we make the following observations:
%
%
\begin{itemize}
    \item The proposed method shows consistent performance in the low training sample regime for both datasets.
    \item For the MNIST dataset, our method outperforms the classical ML techniques - SVM, KNN, and DT. When compared with DL ($2$-layers) the method closely matches the accuracy up to $7$ training samples/class.
    \item For the Iris dataset, our method has the best performance when trained with $2$ samples/class. DL ($2$-layers) gave the least accuracy throughout the low training sample regime.
\end{itemize}  
\section{Conclusions and Future Research Directions}
As evident from the results, TT based classification gives a consistent performance in low sample regime compared to classical ML/DL techniques. This method can be particularly useful when the number of available training samples is less. As the size of training data increases, conventional ML/DL methods outperform our method. A direction for future research is to investigate whether we can combine TT based method with ML/DL methods to yield a superior hybrid algorithm. 

A significant advantage of the proposed method is that it need not be re-trained from scratch if a new class (with new data samples) is added. The {\it representation} vectors of all the existing classes will not change. Only the representation vector for the new class needs to be computed. In contrast, such a scenario would require a complete re-designing and re-training in the case of ML and DL.

Our method has fewer hyperparameters than ML/DL methods. It can be noticed that we have not performed hyperparameter tuning in this work since we are dealing with very few training samples. Future work could involve using multiple chaotic maps (such as logistic map) and also designing a network with multiple layers of chaotic neurons. We expect that such modifications can further increase the accuracy. 

To conclude, we have for the first time proposed a novel chaos based neural architecture which makes use of the property of {\it topological transitivity}. In our architecture, the non-linearity and chaos is intrinsic to the neuron unlike conventional ANN. Earlier research (\cite{rat_place_units} and~\cite{hippocampus_spatial}) highlight the presence of neurons in the hippocampus (of the rat's brain) which are sensitive to a particular point in space. In a similar vein, our method proposes {\it temporally} sensitive neurons. In the proposed model, the {\it firing time} of the chaotic GLS neuron required to match the response of the stimulus is a discriminating feature to distinguish different classes. Thus, our research is an initial step towards employing chaos (and its fascinating properties) in an intrinsic fashion to design novel learning architectures and algorithms that are inspired from the biological brain.
\section*{Acknowledgment}
H.N.B. thanks ``The University of Trans-Disciplinary Health Sciences and Technology (TDU)'' for permitting this research as part of the PhD program. The authors gratefully acknowledge the financial support of Tata Trusts. 
\bibliographystyle{unsrt}

\end{document}